\documentclass[12pt]{article}
\pdfoutput=1
\usepackage[utf8]{inputenc}
\usepackage{indentfirst}
\usepackage{amsmath,amsthm,amssymb}
\usepackage{graphicx}
\usepackage{setspace}
\usepackage{hyperref}
\usepackage[all]{hypcap}

\newcommand*{\I}{\rm{i}} 

\title{The dilute Temperley--Lieb O($n=1$) loop model on a semi infinite strip: 
the sum rule}

\author{A. Garbali$^{1,2,3}$ and B. Nienhuis$^{2}$
\vspace{0.5cm}
\\
$^{1}$LPTHE, CNRS UMR 7589, \\ 
Universit\'e Pierre et Marie Curie (Paris 6),\\
4 place Jussieu, 75252 Paris cedex 05, France.
\and
$^{2}$Instituut voor Theoretische Fysica,\\
Universiteit van Amsterdam,
Science Park 904, \\
1098 XH Amsterdam, The Netherlands.
\and
$^{3}$ARC Centre of Excellence for Mathematical and Statistical Frontiers (ACEMS), \\School of Mathematics and Statistics, University of Melbourne, \\Parkville, Victoria 3010, Australia.}
\date{}

\begin{document}

\maketitle
\begin{abstract}
This is the second part of our study of the ground state eigenvector of the transfer matrix of
the dilute Temperley--Lieb loop model with the loop weight $n=1$ on 
a semi infinite strip of width $L$ \cite{GN}. We focus here on
the computation of the normalization (otherwise called the sum rule) $Z_L$ of the ground state
eigenvector, which is also the partition function of the critical site percolation model.
The normalization $Z_L$ is a symmetric polynomial in the inhomogeneities
of the lattice $z_1,..,z_L$. This polynomial satisfies several recurrence relations 
which we solve independently in terms of Jacobi--Trudi like determinants. Thus we provide a
few determinant expressions for the normalization $Z_L$.
\end{abstract}

\section{Introduction}\label{sec1}
The inhomogeneous loop models on the two dimensional lattices on semi-infinite domains 
 with periodic and open boundary conditions were actively studied in the last decade. This concerns, in particular, 
the cases when the loop weight $n=1$\footnote{In fact we are interested in the regime when the crossing parameter $q$, which is related to $n$ by $n=-q^{-4}-q^4$, is a certain root of unity. More precisely, let $\omega=e^{\I\pi/3}$, from now on we assume that $q$ is specialized as $q=\omega$.}, which we assume everywhere below. Most famous 
examples of these models are: the Temperley--Lieb (TL) loop model 
\cite{BGN,RS,Ca,MitraNienhuis,MitraNienhuis2,PDFPZJ,PDF1,IP,dGPS,dGNP,PZJ},
the Brauer loop model (BL) \cite{N_1993,MNR_98,GN_2005,PDFPZJ1,AKPZJ,PDF1,APZJ} and recently the dilute 
Temperley--Lieb (dTL) loop model \cite{PDF,GN,FN}.
These models have many connections to combinatorics, critical percolation,
geometric representation theory, etc. These connections are discussed in more details in the given references.

We are interested in studying the ground state eigenvector
of the transfer matrix, its normalization and correlation functions.  
Thanks to the fact that the ground state has polynomial entries it was possible to develop 
a procedure to compute these entries
using some $q$-difference equations (loosely called the quantum Knizhnik--Zamolodchikov ($q$KZ) equations) and certain recurrence relations. This approach is analogous to the procedure developed by Di Francesco and Zinn-Justin for the TL model at $n=1$. 
We have done this computation for the dTL model with open boundary conditions 
in \cite{GN}. In the present work we
calculate the normalization $Z_L$ of the ground state $\Psi_L$
of the dTL model with open boundaries.

Similarly as in the other loop models (TL and BL) $Z_L$ is a symmetric polynomial in the inhomogeneity parameters $z_1,..,z_L$ which obeys certain recurrence relations. The first recurrence relation is related to a factorization of the $R$ matrix into two operators. One of these operators gives rise to a map $\Psi_L$ to $\Psi_{L-1}$. The normalization of $\Psi_L$ in this case satisfies the recurrence:
\begin{align}\label{rec1}
Z_L(z_1,..,z_{L-1}=z_{L-1}\omega,z_L=z_{L-1}/\omega)=
F(z_1,..,z_{L-2}|z_{L-1})Z_{L-1}(z_1,..,z_{L-1}).
\end{align}
This recurrence relation fixes $Z_L$ completely once the initial condition is specified. 
The computation of the polynomial\footnote{Strictly speaking $F$ in (\ref{rec1}) as well as the function $P$ (\ref{rec2}) are rational functions. However, they have a trivial denominator, hence we call them polynomials in what follows.} $F$ is a result of our previous work \cite{GN}, we will
specify it later.
The second recurrence relation, which is unrelated to (\ref{rec1}) has the form:
\begin{align}\label{rec2}
Z_L(z_1,..,z_{L-1}=z_{L-1},z_L=-z_{L-1})=
P(z_1,..,z_{L-2}|z_{L-1})Z_{L-2}(z_1,..,z_{L-2}),
\end{align}
where the polynomial $P$ will be given later. We expect that this recurrence relation is coming from another factorization of the $R$-matrix\footnote{Our expectation is based on the study of the $U_q(A_2^{(2)})$ ``spin'' model whose $R$-matrix has two different factorizations \cite{Ga}. The algebra $U_q(A_2^{(2)})$ defines the 
Izergin--Korepin vertex model \cite{IK} which is related to the dTL model by a certain basis transformation.}, however, we do not prove this in the current work. This recurrence relation has a unique solution given the initial condition. 

The same type of recurrence relations appear in
the dTL model with periodic boundary conditions \cite{PDF}. Let $F^p$ and $P^p$ be some polynomials in the variables $z_1,..,z_{L-1}$, where the superscript $p$ refers to the periodic boundary conditions. Let also $Z_L^p$ be the
normalization of the ground state vector of the periodic transfer matrix of the dTL model, then
\begin{align}\label{rec1P}
Z^p_L(z_1,..,z_{L-1}=z_{L-1}\omega,z_L=z_{L-1}/\omega)=
F^p(z_1,..,z_{L-2}|z_{L-1})Z^p_{L-1}(z_1,..,z_{L-1}).
\end{align}
\begin{align}\label{rec2P}
Z^p_L(z_1,..,z_{L-1}=z_{L-1},z_L=-z_{L-1})=
P^p(z_1,..,z_{L-2}|z_{L-1})Z^p_{L-2}(z_1,..,z_{L-2}),
\end{align}
In this case (\ref{rec1P}) is solved by a determinant of elementary symmetric polynomials 
\cite{PDF} which can be identified with a skew Schur function of certain partition of the staircase shape.
We are going to use this solution to find a determinant expression solving
(\ref{rec1}). We will also show how to compute $Z_L$ and $Z_L^p$ using
the second recurrence relations (\ref{rec2}) and (\ref{rec2P}), respectively. The latter also give  
determinant expressions for $Z_L$ and $Z_L^p$. This time the matrix entries of the determinants are expressed using a different set of symmetric polynomials. 
Therefore, the two equations (\ref{rec1}) and (\ref{rec2}) (as well as (\ref{rec1P}) and (\ref{rec2P}) in the preiodic case) lead us to two different determinant representations of $Z_L$ ($Z_L^p$ in the periodic case)
which are computed independently. These representations must be related by some transformation which is unknown to us. 

An interesting observation in the course of our computations was
to realize that the polynomial $P^p(z_1,..,z_L|\zeta)$ is the Baxter's $Q$-function 
\cite{Baxter} of the (conjectural) ground state of the 
corresponding spin chain, which is the  $U_q(A_2^{(2)})$ integrable 
model (Izergin--Korepin model \cite{IK}). This means that the roots of this polynomial, regarded 
as a polynomial in $\zeta$, are the Bethe
roots of the IK spin chain. In particular, it allowed us to compute this state for small systems and compare to the $q$KZ-based calculation of our previous work.
The comparison of these results is possible because 
the ground states of the loop model and the one of the vertex model are related by a linear 
transformation \cite{Nienhuis1}. For small systems ($L\leq 5$) it is possible to match them 
completely. 
The details of this calculation on the IK model side appear in \cite{Ga}.

The outline of the paper is as follows. 
We will start by introducing the dTL model with $n=1$ in Section \ref{sec2}. For a more
detailed introduction we refer to \cite{GN}. In Section \ref{sec3} we will
show how to solve (\ref{rec1}) for $Z_L$ using the known solution $Z_L^p$ of (\ref{rec1P}) computed in \cite{PDF}. 
In Section \ref{sec4} we show how to solve (\ref{rec2P}) and (\ref{rec2}).
The conclusion is given in Section \ref{sec5}. 

\section{The model}\label{sec2}
\begin{figure}[htb]
\centering
\includegraphics[width=0.7\textwidth]{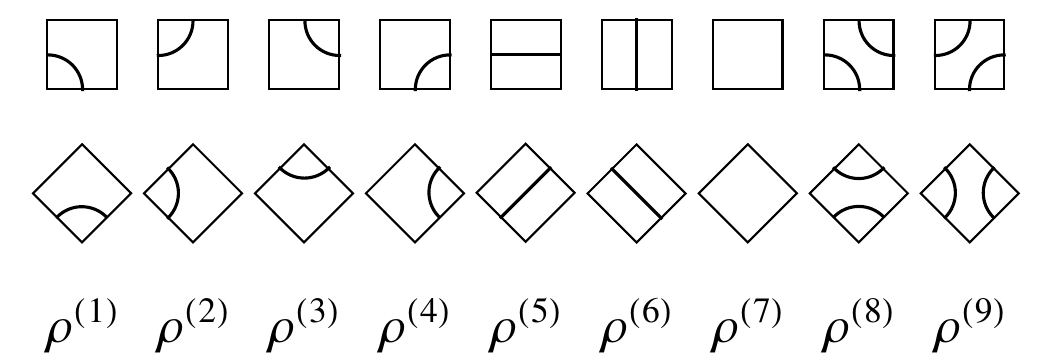}
\caption{The bulk plaquettes (first row). The graphical representation of the 
nine operators (second row) acting on the link patterns (see Fig. \ref{figaction}). 
Graphically these are the 45 degrees tilted versions of the bulk plaquettes, they are called
$\rho^{(1)}$, $\rho^{(2)},..,\rho^{(9)}$ respectively.}
\label{figplaq}
\end{figure}

\begin{figure}[htb]
\centering
\includegraphics[width=0.4\textwidth]{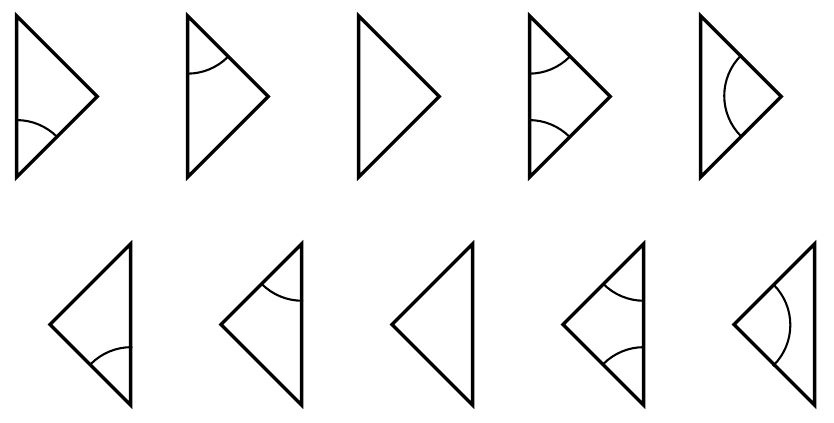}
\caption{The left (top row) and the right (bottom row) boundary plaquettes. The corresponding left boundary
operators will be called $\kappa_l^{(1)},..,\kappa_l^{(5)}$ and the right boundary operators
$\kappa_r^{(1)},..,\kappa_r^{(5)}$, respectively.}
\label{figbplaq}
\end{figure}

\begin{figure}[htb]
\centering
\includegraphics[width=0.35\textwidth]{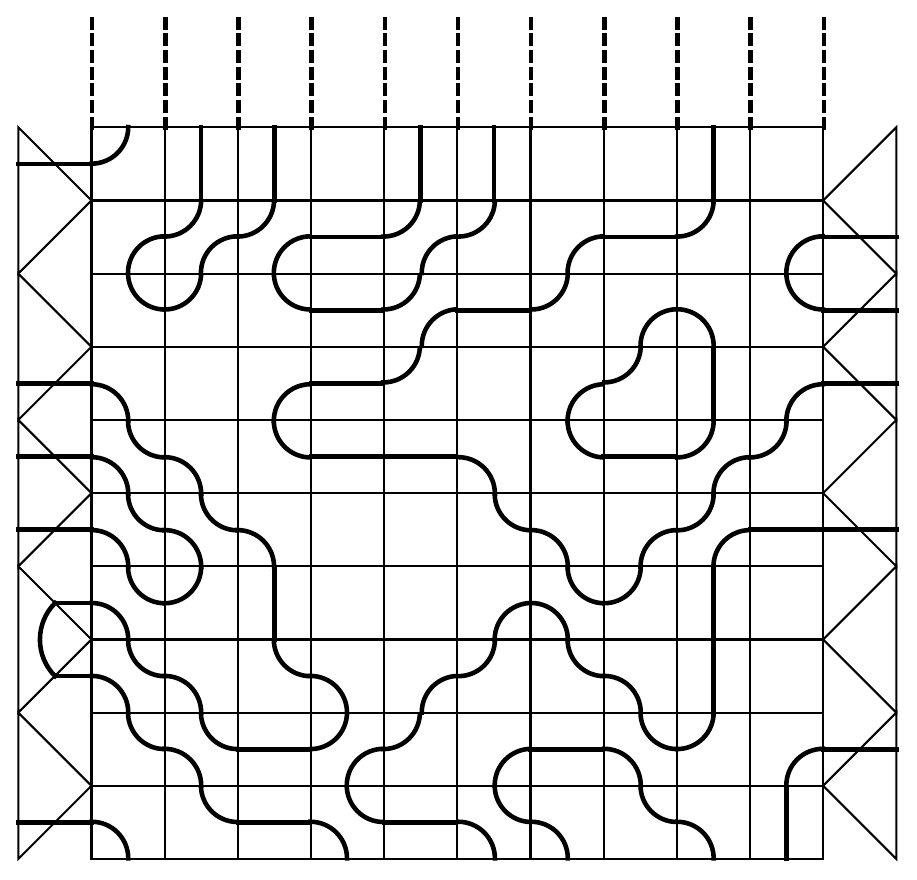}
\caption{A typical configuration of the dilute O$(n)$ loop model on a fragment of the
semi-infinite strip.}
\label{figlp}
\end{figure}

The dTL loop model is defined on the square lattice by decorating the faces of the lattice
with one of the nine plaquettes (Fig.  \ref{figplaq})
in such a way that all loops in the bulk are continuous. 
The loops may end on the boundaries or form closed cycles in the bulk. We consider the
model on a semi-infinite strip which is finite in the horizontal direction 
and infinite in the vertical. If we identify the two vertical boundaries of the strip 
then such boundary conditions are called periodic. If we forbid loops to end at the
vertical boundaries then we need to include two boundary plaquettes (the third and fifth 
on Fig.  \ref{figbplaq}). These boundary conditions are called closed or reflecting.
If we allow the loops to end at the vertical boundaries, as on Fig.  \ref{figlp}, 
then it gives rise to open boundary conditions. The latter case requires to consider three more
boundary plaquettes along with the two of the reflecting case. All five boundary plaquettes are
presented on Fig.  \ref{figbplaq}. The dTL model with open boundary conditions is
the one we study here. We also shortly discuss and present a result for the periodic dTL model.

\begin{figure}[htb]
\centering
\includegraphics[width=1\textwidth]{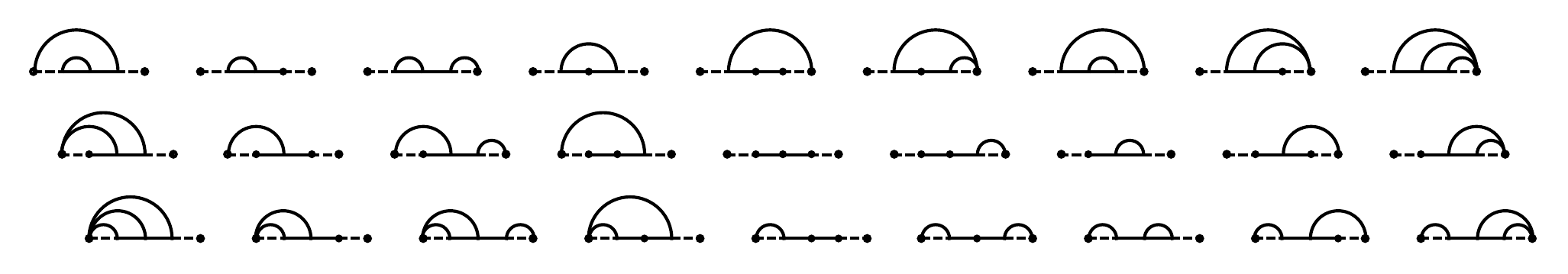}
\caption{The basis elements of LP$_3$.}
\label{figlp3}
\end{figure}

The operators $\rho^{(i)}$ as well as $\kappa_l^{(i)}$ and $\kappa_r^{(i)}$ naturally act
in the space of link patterns LP$_L$.
The space LP$_L$ is spanned by all possible connectivities of $L + 2$ vertices on a straight
horizontal line with certain restrictions. The first and the last vertex are called the
boundary vertices, while the vertices in between are called the bulk vertices.
A bulk vertex can be disconnected from any other vertex (thus called unoccupied)
or connected (occupied) only once to another bulk or boundary vertex. A boundary vertex can
be disconnected from the other vertices or connected to any number of distinct bulk
vertices (not the other boundary vertex). We also require that there are no crossings
in the connectivity. For $L = 3$ all possible connectivities, or the basis elements of LP$_3$,
are depicted on Fig. \ref{figlp3}.
\begin{figure}[htb]
\centering
\includegraphics[width=0.4\textwidth]{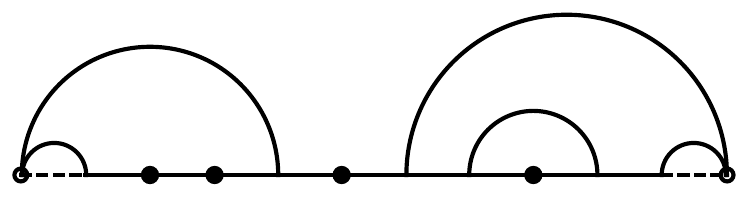}
\caption{The link pattern that corresponds to the configuration on Fig. \ref{figlp}.}
\label{figlinkp}
\end{figure}
The space LP$_L$ is in one to one correspondence with $\{-1,0,1\}^{\otimes L}$ since to each site in the link pattern we can assign $-1,0$ and $1$ if this site is linked to the left, empty and linked to the right, respectively. 
Every configuration of the loop model corresponds to a link pattern $\pi$. This can be seen
by erasing all closed loops in the bulk of the strip and all links connecting two
vertical boundary points. The configuration on Fig. \ref{figlp} corresponds to the
link pattern shown on Fig. \ref{figlinkp}.

The object of interest is the ground state vector $\Psi_L$ of the transfer matrix, which
we will introduce below, can be
represented as a vector in the space of link patterns:
\begin{align}\label{Psieq}
\Psi_L =\sum_{\pi \in \text{LP}_L}\psi_{\pi}|\pi\rangle.
\end{align}
The computation of its components $\psi_{\pi}$ was the subject of our previous work \cite{GN}.

\begin{figure}[htb]
\centering
\includegraphics[width=0.7\textwidth]{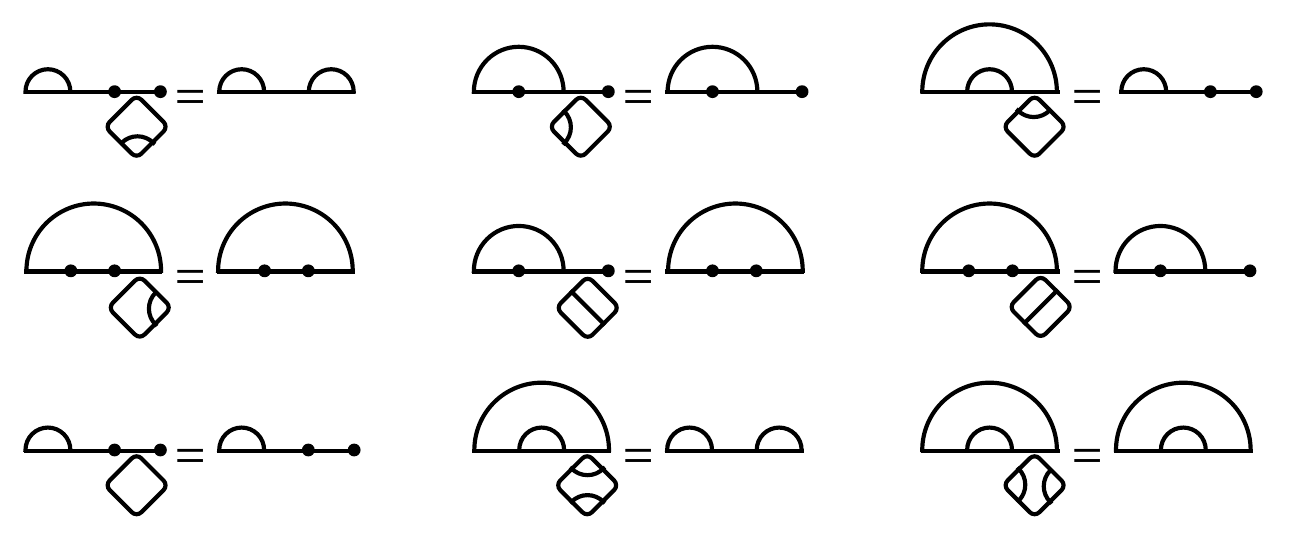}
\caption{Several  examples of the action of the operators $\rho^{(i)}$ .}
\label{figaction}
\end{figure}

\begin{figure}[htb]
\centering
\includegraphics[width=0.8\textwidth]{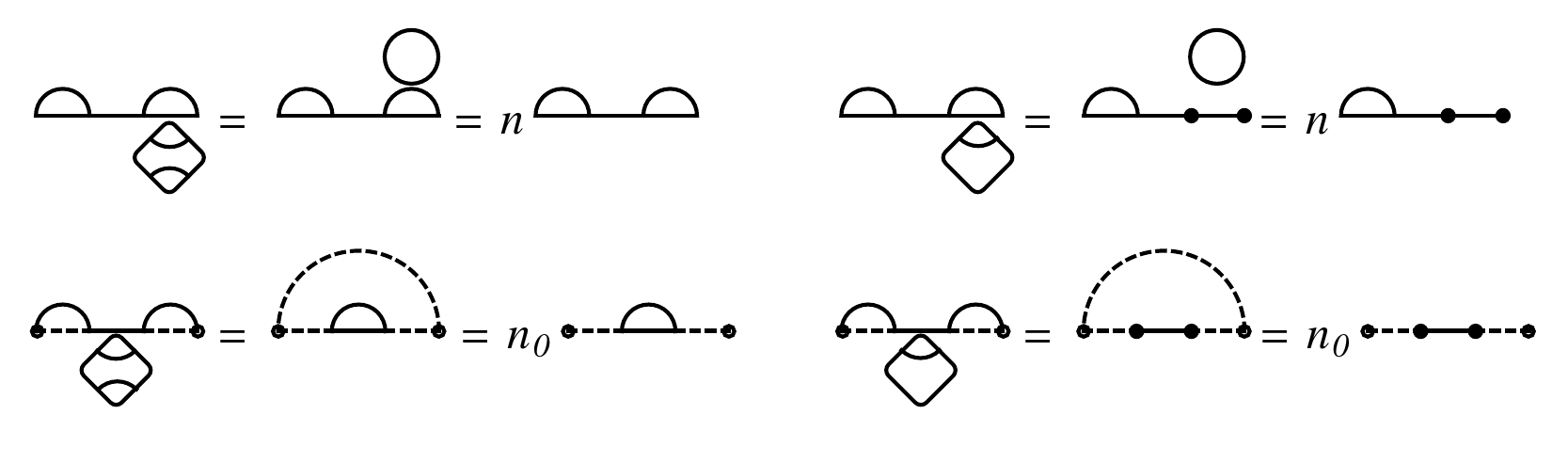}
\caption{The operators $\rho^{(3)}$ and $\rho^{(8)}$ produce a closed loop
or a line connecting two vertical boundaries which is represented by the dashed
semi-circle. Both, the loop weight $n$ and the weight of the boundary to boundary line $n_0$, we set to $1$.}
\label{figaction1}
\end{figure}

\begin{figure}[htb]
\centering
\includegraphics[width=0.4\textwidth]{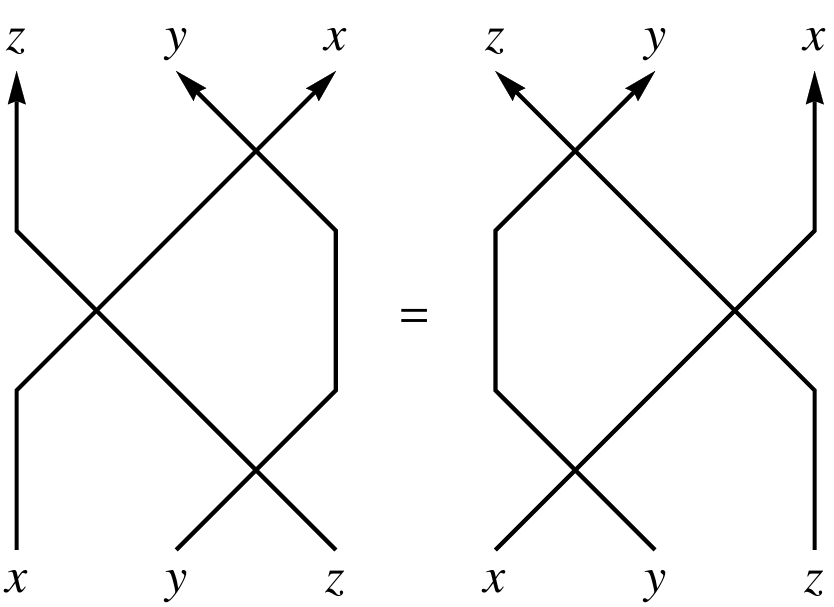}
\caption{The Yang--Baxter equation.}
\label{figYB}
\end{figure}

The action of the bulk operators $\rho^{(i)}$ on the link patterns goes as follows.
An operator $\rho^{(i)}_j$, supplied with the position index $j$,  acts non trivially on two neighbouring vertices $j$ and $j+1$ of a link pattern
if the occupancy of the vertices $j$ and $j+1$ coincides with the occupancy of the north
west (NW) edge and the north east (NE) edge of $\rho^{(i)}$, respectively.
Then we need to connect the middle of the NW edge of $\rho^{(i)}$ with the $j$-th
vertex of the link pattern and the middle of the NE edge of $\rho^{(i)}$
with the $j+1$ vertex of the link pattern. In the resulting link pattern the connectivity at
the points $j$ and $j+1$ will be that of the middle points of the south west edge and south
east edge of the operator $\rho^{(i)}$. A few examples are presented on Fig. \ref{figaction}
and Fig. \ref{figaction1}.
The boundary plaquettes act on the first and the last bulk points of link patterns in a
similar way. A few examples of this action are presented on Fig. \ref{figaction}.
\begin{figure}[htb]
\centering
\includegraphics[width=1\textwidth]{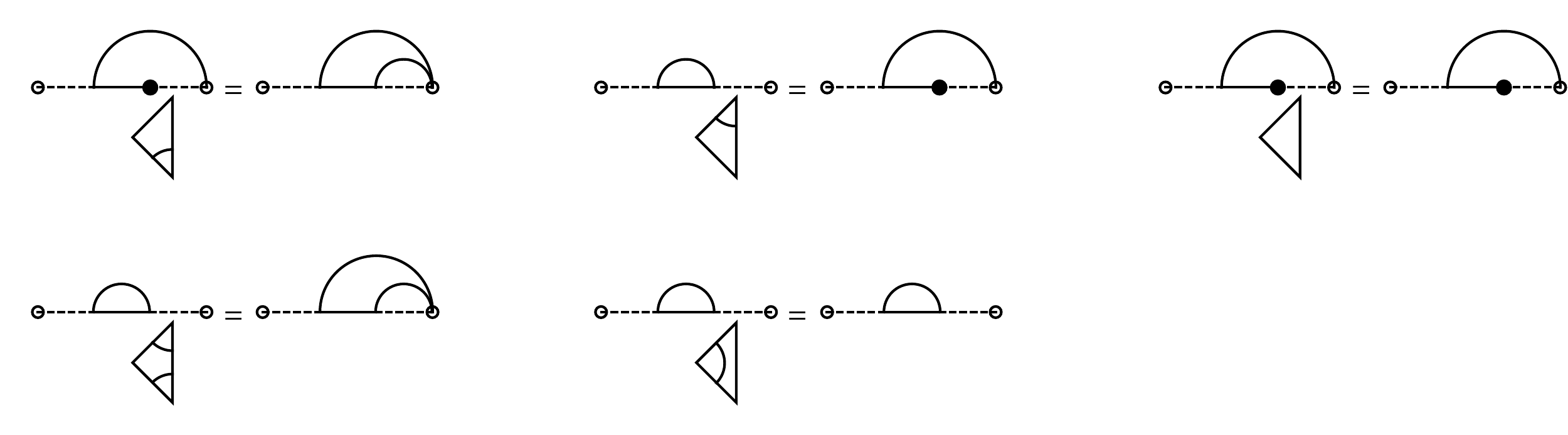}
\caption{The action of the $\kappa_r$-operators.}
\label{figbaction}
\end{figure}
Now we need to define the $\check{R}$-matrix, $R$-matrix, the $K$-matrices and then the
transfer matrix of the dTL model. The $\check{R}$-matrix is the weighted action of the operators $\rho^{(i)}$ represented by the second row on Fig. \ref{figplaq}:
\begin{align}\label{Rmx}
\check{R}_j(z_j,z_{j+1})=\sum_{i=1}^{9}\rho_j^{(i)}r_i(z_j,z_{j+1}).
\end{align}
\begin{figure}[htb]
\centering
\includegraphics[width=0.4\textwidth]{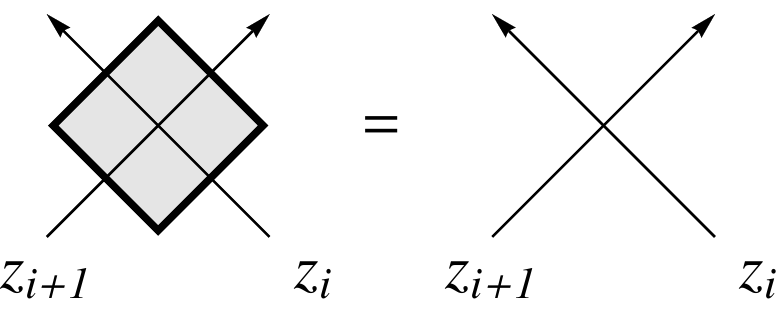}
\caption{The $\check{R}_i(z_i,z_{i+1})$-matrix.}
\label{figRcheck}
\end{figure}
\begin{figure}[htb]
\centering
\includegraphics[width=0.45\textwidth]{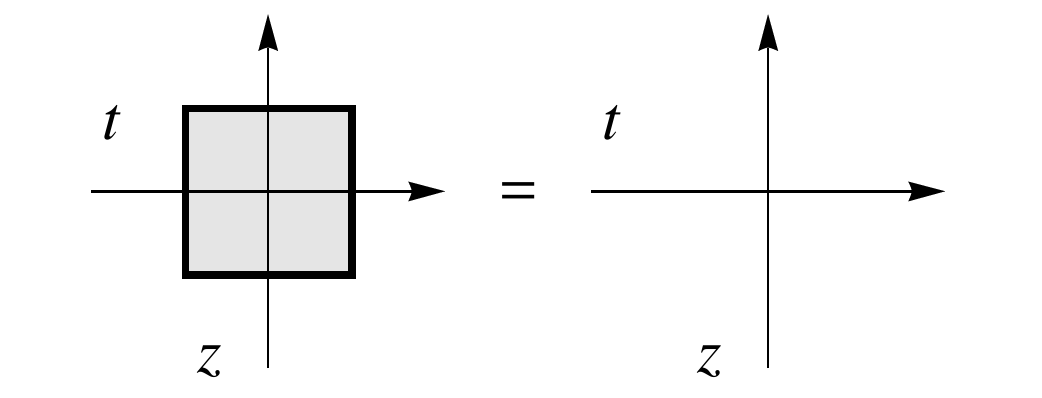}
\caption{The $R(z,t)$-matrix.}
\label{figR}
\end{figure}
Since the $\check{R}$-matrix acts on two points of the vector space of link patterns, it carries two
rapidity parameters $z_j$ and $z_{j+1}$. On Fig. \ref{figRcheck} we show the graphical
representation of the $\check{R}$-matrix, where the spectral parameters are carried
by the straight oriented lines. To obtain the $R$-matrix we simply take the $\check{R}$-matrix
and rotate it by $45$ degrees clockwise, 
see Fig. \ref{figR}.
The integrable $\check{R}$-matrix (as well as $R$) depends on the ratio of two rapidities,
so $\check{R}_i(z_i,z_{i+1})\propto \check{R}_i (z_{i}/z_{i+1})$. The integrability
requires that it satisfies the Yang--Baxter (YB) equation \cite{Baxter}
\begin{align}\label{YB}
\check{R}_{i+1}(z/y)\check{R}_{i}(z/x)\check{R}_{i+1}(y/x)=
\check{R}_{i}(y/x)\check{R}_{i+1}(z/x)\check{R}_{i}(z/y).
\end{align}
Graphically it is shown on Fig. \ref{figYB}.
This equation defines the integrable weights of the $\check{R}$-matrix
\begin{align}\label{Rweights1}
&r_1(z)=r_2(z)=r_3(z)=r_4(z)=\omega  (\omega +1) z,~~~
r_5(z)=r_6(z)=r_7(z)=z^2-1, \nonumber \\
&r_8(z)=-\left(\omega +z\right) \left(\omega ^2 z+1 \right),
~~~ r_9(z)=\left(\omega ^2+z\right) \left(\omega  z+1 \right).
\end{align}
Here $\omega$ is a root of $-1$, $\omega=e^{\I\pi/3}$, which implies the condition on
the loop weight $n = 1$. The dTL loop model with generic value of $n$ was obtained in \cite{Nienhuis1,Nienhuis2}. 

\begin{figure}[htb]
\centering
\includegraphics[width=0.5\textwidth]{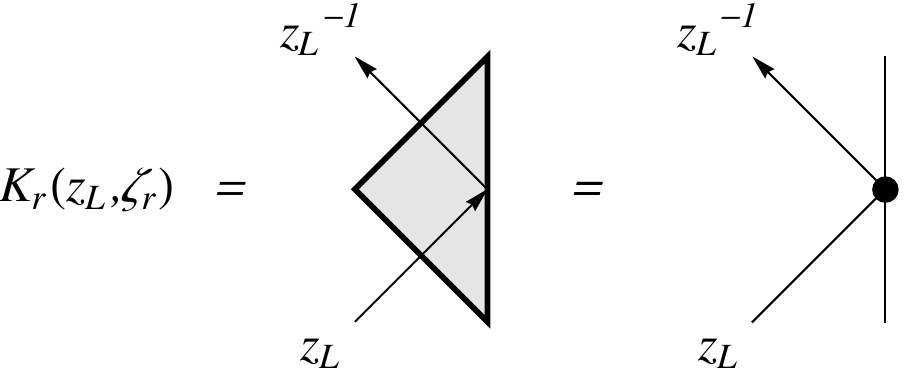}
\caption{The operator $K_r(z_L,x_r)$.}
\label{figKmx2}
\end{figure}

\begin{figure}[htb]
\centering
\includegraphics[width=0.35\textwidth]{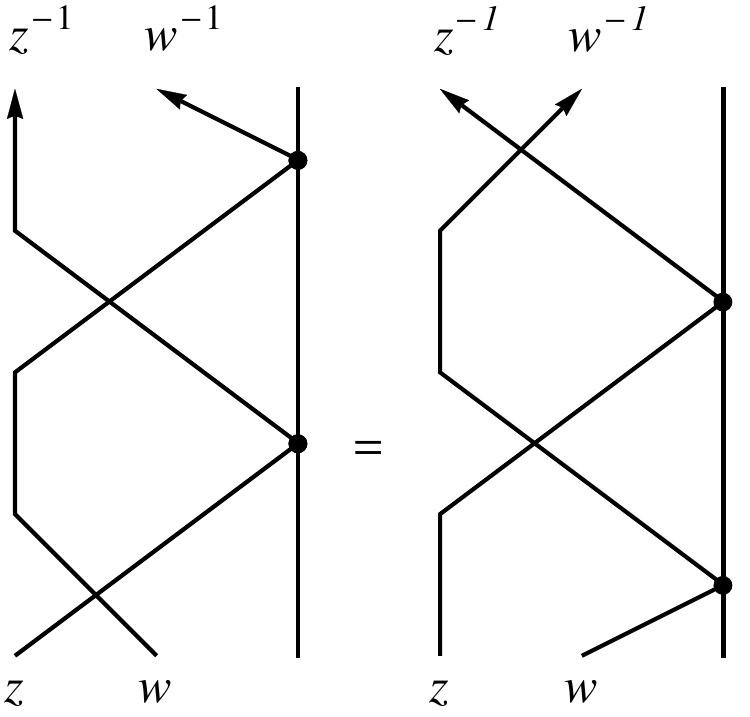}
\caption{The boundary Yang--Baxter equation.}
\label{figBYB}
\end{figure}

\begin{figure}[htb]
\centering
\includegraphics[width=0.7\textwidth]{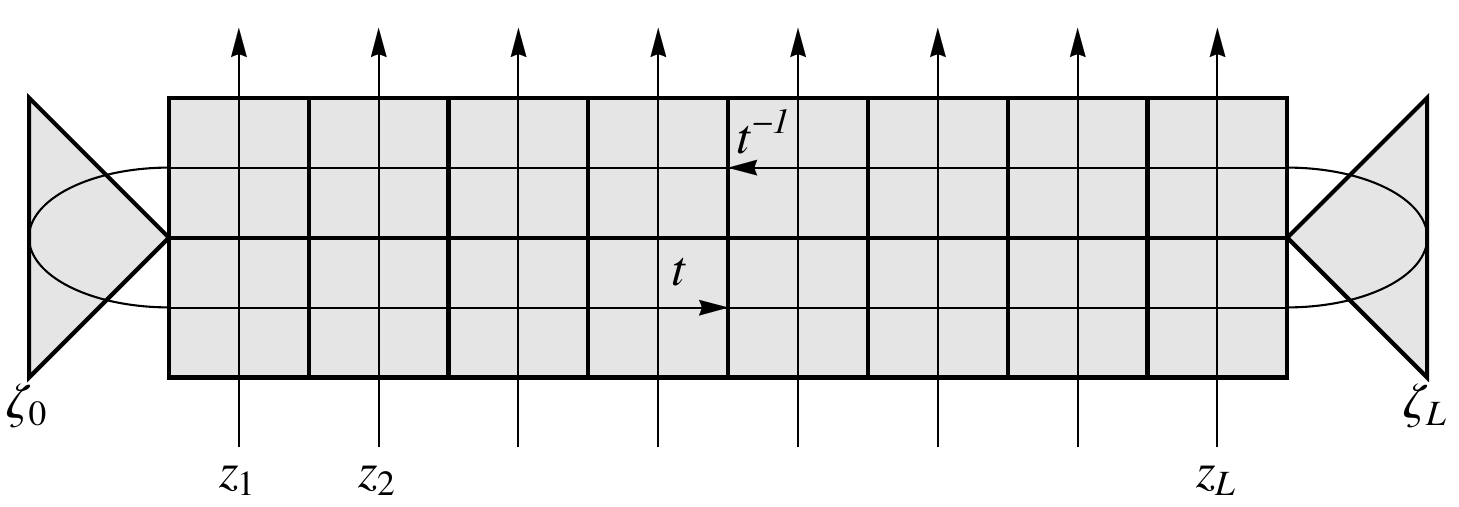}
\caption{The graphical representation of the transfer operator.}
\label{figTmx}
\end{figure}

The $K$-matrix is a combination of the five boundary plaquettes. There is the left
$K$-matrix $K_l$ and the right $K$-matrix $K_r$
\begin{align}\label{Kmx}
K_l(z_1,x_l)=\sum_{i=1}^{5}\kappa^{(i)}_l k_{i,l}(z_1,x_l),~~~
K_r(z_L,x_r)=\sum_{i=1}^{5}\kappa^{(i)}_r k_{i,r}(z_L,x_r)
\end{align}
Here, $x_l$ and $x_r$ play the role of the boundary rapidities. Note, in our previous work \cite{GN} we used the parameters 
$\zeta_l,\zeta_r$, which are related to $x_l,x_r$ by $\zeta_i=\omega x_i(x_i^2+1)^{-1}$. The $K$-matrices also have a 
convenient graphical representation, as shown on Fig. \ref{figKmx2}.
The $R$ and $K$-matrices should satisfy the Sklyanin's reflection equation \cite{Skl} 
also called the boundary  Yang--Baxter equation (BYB). For the right
boundary it reads
\begin{align}\label{BYB}
\check{R}_{L-1}(w/z)K_{r}(z,x_r)\check{R}_{L-1}(1/(w z))K_{r}(w,x_r)=
K_r(w,x_r)\check{R}_{L-1}(1/(w z))K_{r}(z,x_r)\check{R}_{L-1}(w/z).
\end{align}
and graphically is presented on Fig. \ref{figBYB}. The graphical representation of the 
left $K$-matrix as well as the corresponding reflection equation are similar to the ones 
of the right $K$-matrix \cite{FN}.
Solving the left boundary reflection equation one obtains
\begin{align}\label{Kweights1}
&k_{1,l}(z,x_l)=k_{2,l}(z,x_l)=-\frac{(\omega+1)(x_l^2+1) \left(z^2-1\right)}{x z},\nonumber \\ 
&k_{3,l}(z,x_l)=\frac{x_l^4 z^2-x_l^2 z^4+3 x_l^2 z^2-x_l^2+z^2}{\omega x_l^2 z^2},\nonumber \\
&k_{4,l}(z,x_l)=-\frac{  (\omega+1)(x_l^2+1) \left(z^2-1\right) 
\left(\omega-z^2\right)}{x_l z^2},\nonumber \\ 
&k_{5,l}(z,x_l)=\frac{-\omega z^4 x_l^2+\omega x_l^2+z^4 x_l^2+z^2 x_l^4+z^2}{\omega z^2 x_l^2}.
\end{align}
The weights of the right boundary $K$-matrix are given by $k_{i,r}(z,x)=k_{i,l}(1/z,x)$, which 
can be achieved by solving the right boundary reflection equation. 
Following the general prescription \cite{Skl} we construct the double row transfer matrix 
(Fig. \ref{figTmx}) using the $R$ and $K$-matrices
\begin{align}\label{Tmx}
T(t|z_1,..,z_L;x_l,x_r)=\text{Tr}\big{(}
R_1(z_1/t)..R_L(z_L/t)K_r(t,x_r)R_L(1/(t z_L))..R_1(1/(t z_1))K_l(t^{-1},x_l)\big{)},
\end{align}
where the trace means that the lower edge of the $K_l(t^{-1},x_l)$ needs to be identified
with the left edge of $R_1(t,z_1)$. The $T$-matrix above is the inhomogeneous transfer matrix,
it depends on the bulk spectral parameters $z_1,..,z_L$ associated to each space of the lattice
and also on the two boundary parameters $x_l$ and $x_r$ associated to the left and the
right boundaries. Due to the YB and the BYB two transfer matrices with different values of
$t$ commute \cite{Skl}
\begin{align}\label{TT}
[T(t_1),T(t_2)]=0.
\end{align}
Therefore the eigenvectors of this transfer matrix must depend on $\{z_1,..,z_L,x_l,x_r\}$,
but not on the parameter $t$.
We also have the following commutation of the $T$-matrix with the $\check{R}$-matrix
and the $K$-matrices
\begin{align}
&\check{R}_i (z_i,z_{i+1})T(t|z_1,..,z_i,z_{i+1},..,z_L;x_l,x_r)=
T(t|z_1,..,z_{i+1},z_i,..,z_L;x_l,x_r)\check{R}_i (z_i,z_{i+1}),,\label{RT} \\
&K_l (z_1,x_l)T(t|z_1,z_2,..,z_L;x_l,x_r)=
T(t|1/z_1,z_2,..,z_L;x_l,x_r)K_l (z_1,x_l),\label{KTl}\\
&K_r (z_L,x_r)T(t|z_1,..,z_{L-1},z_L;x_l,x_r)=
T(t|z_1,...z_{L-1},1/z_L;x_l,x_r)K_r (z_L,x_r).\label{KTr}
\end{align}
In the previous work we were focused on finding the highest eigenvector $\Psi_L$ of the
transfer matrix $T_L$. If we properly normalize the $T$-matrix we can write
$T_L \Psi_L = \Psi_L$. Using (\ref{RT}), (\ref{KTl}) and (\ref{KTr}) we
find the $q$KZ equations:
\begin{align}
&\check{R}(z_i,z_{i+1})\Psi_L (z_1,..,z_i,z_{i+1},..,z_L;x_l,x_r)=
W(z_i,z_{i+1})\Psi_L (z_1,..,z_{i+1},z_{i},..,z_L;x_l,x_r),\label{R$q$KZ} \\
&K_l(z_1,x_l)\Psi_L (z_1,..,z_L;x_l,x_r)=
U_l(z_1,x_l)\Psi_L (1/z_1,..,z_L;x_l,x_r),\label{B$q$KZl} \\
&K_r(z_L,x_r)\Psi_L (z_1,..,z_L;x_l,x_r)=
U_r(z_L,x_r)\Psi_L (z_1,..,1/z_L;x_l,x_r),\label{B$q$KZr}
\end{align}
where $W(z_i,z_{i+1})$, $U_l(z_1,x_l)$ and $U_r(z_L,x_r)$ are the normalizations
of the $\check{R}$-matrix, $K_l$-matrix and $K_r$-matrix respectively. They can be written 
as combinations of weights of $\check{R}$ and $K_l$ and $K_r$, respectively, as:
\begin{align}
&W(z_i,z_{i+1})=r_2(z_i,z_{i+1})+r_6(z_i,z_{i+1}), \nonumber \\
&U_l(z_1,x_l)=k_{1,l}(z_1,x_l)+k_{3,l}(z_1,x_l)=
k_{2,l}(z_1,x_l)+k_{4,l}(z_1,x_l)+k_{5,l}(z_1,x_l), \nonumber \\
&U_r(z_L,x_r)=k_{1,r}(z_L,x_r)+k_{3,r}(z_L,x_r)=
k_{2,r}(z_L,x_r)+k_{4,r}(z_L,x_r)+k_{5,r}(z_L,x_r).\nonumber
\end{align}
We used in our last work
 (\ref{R$q$KZ})-(\ref{B$q$KZr}) in order to compute the components $\psi_{\pi}$ of the vector $\Psi_L$.
This computation, however, is not possible without the recurrence relation which we will
consider in the following section.

\section{The first recurrence relation}\label{sec3}
\begin{figure}[htb]
\centering
\includegraphics[width=0.8\textwidth]{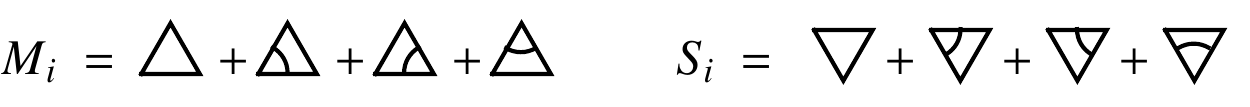}
\caption{The $M_i$ and $S_i$ operators.}
\label{figMS}
\end{figure}

\begin{figure}[htb]
\centering
\includegraphics[width=0.5\textwidth]{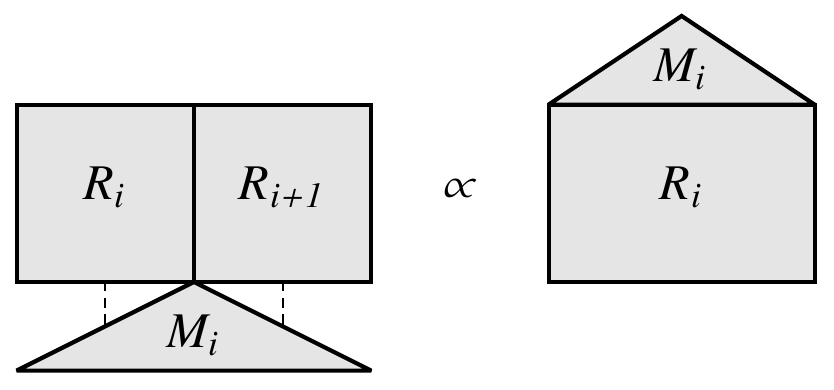}
\caption{The graphical version of (\ref{MRR}). The $M$-operator is represented 
by the triangle.}
\label{figYenB}
\end{figure}

In this section we will discuss the first recurrence relation (\ref{rec1}) for the
normalization $Z_L$ of the ground state vector of the transfer matrix. It is defined as the sum of all components of $\Psi_L$
\begin{align}\label{norm}
Z_L(z_1,..,z_L;x_l,x_r)=\sum_{\pi \in \text{LP}_L} \psi_{\pi}(z_1,..,z_L;x_l,x_r),
\end{align}
The derivation of this recurrence relation was given in our
previous work. It follows from a factorization property of the $R$-matrix at a special value of its parameter. More precisely $\check{R}(z_i \omega, z_i/\omega)$
factorizes into two operators
\begin{align}\label{RMS}
\check{R_i}(z\omega,z/\omega)=(\omega^2+\omega)z^2 S_i M_i.
\end{align}
This gives rise to a ``modified'' version of the YB equation. It
involves two $R$-operators and one $M$-operator:
\begin{align}\label{MRR}
&M_i\check{R_i}(z_i \omega)R_{i+1}(z_i/\omega) =(z_i^2-1) R_i(t,z_i)M_i,\nonumber \\
&M_i\check{R_i}(z_i/\omega)R_{i+1}(z_i \omega) =(z_i^2-1) R_i(t,z_i)M_i.
\end{align}
In the quantum group literature this is related to the quasi-triangularity condition of 
the corresponding Hopf algebra. The operator
$M$ maps two sites into one site and hence merges the two $R$-matrices in (\ref{MRR}) into one after the
substitution $z_i=z_i \omega$ and $z_{i+1}=z_i/\omega$.
The graphical representation of $M$, $S$ and (\ref{MRR}) are presented on Fig. \ref{figMS} and Fig. \ref{figYenB}.
If we apply $M_i$ to the transfer matrix using (\ref{MRR}) we get:
\begin{align}\label{MT}
M_i T_{L+1}(t|z_1,..,z_i \omega,z_i/\omega,z_{i+1},..,z_L;x_l,x_r)=
T_L(t|z_1,..,z_i,z_{i+1},..,z_L;x_l,x_r)M_i,
\end{align}
assuming that the normalization of the transfer matrix is chosen appropriately.
Applying  (\ref{MT}) to the ground state we find the desired recurrence relation \cite{GN}
\begin{align}\label{MGS}
M_i \Psi_{L+1}(z_1,..,z_i \omega,z_i/\omega,z_{i+1},..,z_{L};x_l,x_r)=
F_i(z_1,..,z_L;x_l, x_r)
\Psi_L(z_1,..,z_i,z_{i+1},..,z_L;x_l,x_r).
\end{align}
The index $i$ in $F_i$ signifies that $F$ has a special dependence on the variable $z_i$.
The explicit form of this polynomial we found in our first paper on the dTL model, it reads:
\begin{align}\label{Ffactor}
F_i(z_1,..,z_L;x_l=z_0,x_r=z_{L+1})=\prod_{0\leq j\neq i\leq L+1}
\frac{(z_i+z_j)(z_i z_j+1)}{z_i z_j}.
\end{align}
Note that we renamed the boundary rapidities $x_l=z_0,~x_r=z_{L+1}$.
The components of $\Psi_{L+1}$ are related to the components of
$\Psi_L$ via (\ref{MGS}). In particular, if $\pi \in$LP$_{L+1}$ has two empty sites at the positions $i$ and
$i+1$, i.e. $\pi=\{\alpha,0,0,\beta\}$ and $\alpha \in \{-1,0,1\}^{i-1}$ and $\beta \in \{-1,0,1\}^{L-i}$,
then the recurrence (\ref{MGS}) maps $\psi_{\{\alpha,0,0,\beta\}}$ at size $L+1$ to
$\psi_{\{\alpha,0,\beta\}}$ at size $L$. When all the sites are empty in $\pi$ we have
\begin{align}\label{eerec}
&\psi_{e}(z_1,..,z_i \omega,z_i/\omega,z_{i+1},..,z_{L};x_l=z_0,x_r=z_{L+1})
=\nonumber\\
&\prod_{0\leq j\neq i\leq L+1}\frac{(z_i+z_j)(z_i z_j+1)}{z_i z_j}
\psi_{e}(z_1,..,z_i,..,z_{L};z_0,z_{L+1}).
\end{align}

If the occupancy in $\pi$ is fixed the sum of all
$\psi_{\pi}$ with such occupancy is equal to $\psi_e$. This is a consequence of the stochasticity of the transfer matrix. Analogous situation happens in the dense TL loop model at $n=1$ \cite{PRGN,dGPS}. 
There are in total $2^L$
different choices of the occupancy for the link patterns in LP$_L$, hence $Z_L=2^L \psi_e$. 
We will omit this constant and simply consider the equation:
\begin{align}
&Z_L(z_1,..,z_i \omega,z_i/\omega,z_{i+1},..,z_{L};z_0,z_{L+1})
=\nonumber\\
&\prod_{0\leq j\neq i\leq L+1}\frac{(z_i+z_j)(z_i z_j+1)}{z_i z_j}
Z_L(z_1,..,z_i,..,z_{L};z_0,z_{L+1}).\label{recZ}
\end{align}

Let us now examine the symmetries of $Z_L$. First of all $Z_L$ is symmetric in $\{z_1,..,z_L\}$.
This can be seen using the $q$KZ equation (\ref{R$q$KZ}) for the components
$\psi_{\alpha,n_i,n_{i+1},\beta}(..,z_i,z_{i+1},..)$ with $\alpha \in \{-1,0,1\}^{i-1}$ and
$\beta \in \{-1,0,1\}^{L-i-1}$ (for more discussions of the $q$KZ equation see \cite{PDFPZJ}, and 
for the dTL \cite{GN}).
Let us consider the following sum:
\begin{align}\label{psum}
\bar{\psi}_{n_i,n_{i+1}}=\sum_{\substack{\alpha \in \{-1,0,1\}^{i-1}\\
\beta \in \{-1,0,1\}^{L-i-1}}}\psi_{\alpha,n_i,n_{i+1},\beta}.
\end{align}
Assuming $W=W(z_i,z_{i+1})$, $r_j=r_j(z_i,z_{i+1})$, $k_{l,j}=k_{l,j}(z_i,z_{i+1})$, 
$\bar{\psi}_{1,-1}=\bar{\psi}_{1,-1}(..,z_{i},z_{i+1},..)$ and
$\tilde{\bar{\psi}}_{1,-1}=\bar{\psi}_{1,-1}(..,z_{i+1},z_i,..)$, the $q$KZ equation
for each combination of $n_i$ and $n_{i+1}$ gives:
\begin{align}
& W \tilde{\bar{\psi}}_{1,-1}=r_9\bar{\psi}_{1,-1}+r_1\bar{\psi}_{0,0}+
r_8(\bar{\psi}_{1,1}+\bar{\psi}_{-1,1}+\bar{\psi}_{1,-1}+\bar{\psi}_{-1,-1}),\nonumber \\
& W \tilde{\bar{\psi}}_{1,1}=r_9\bar{\psi}_{1,1},~~~
W \tilde{\bar{\psi}}_{-1,-1}=r_9\bar{\psi}_{-1,-1},~~~
W \tilde{\bar{\psi}}_{-1,1}=r_9\bar{\psi}_{-1,1},~~~\nonumber \\
& W \tilde{\bar{\psi}}_{0,0}=r_7\bar{\psi}_{0,0}+
r_3(\bar{\psi}_{1,1}+\bar{\psi}_{-1,1}+\bar{\psi}_{1,-1}+\bar{\psi}_{-1,-1}) \nonumber
\end{align}
Since $W=r_3+r_8+r_9=r_2+r_7$ the sum of these
five equations gives:
\begin{align}\label{psum2}
\tilde{\bar{\psi}}_{1,1}+\tilde{\bar{\psi}}_{1,-1}+\tilde{\bar{\psi}}_{-1,1}+
\tilde{\bar{\psi}}_{-1,-1}+\tilde{\bar{\psi}}_{0,0}=\bar{\psi}_{1,1}+
\bar{\psi}_{-1,1}+\bar{\psi}_{1,-1}+\bar{\psi}_{-1,-1}+\bar{\psi}_{0,0}.
\end{align}
If we take now the remaining equations
\begin{align}
& W \tilde{\bar{\psi}}_{0,1}=r_4\bar{\psi}_{0,1}+r_6\bar{\psi}_{1,0},\nonumber \\
& W \tilde{\bar{\psi}}_{1,0}=r_2\bar{\psi}_{1,0}+r_5\bar{\psi}_{0,1},\nonumber \\
& W \tilde{\bar{\psi}}_{0,-1}=r_4\bar{\psi}_{0,-1}+r_6\bar{\psi}_{-1,0},\nonumber \\
& W \tilde{\bar{\psi}}_{-1,0}=r_2\bar{\psi}_{-1,0}+r_5\bar{\psi}_{0,-1},\nonumber
\end{align}
we obtain a similar result:
\begin{align}\label{psum3}
\tilde{\bar{\psi}}_{0,1}+\tilde{\bar{\psi}}_{1,0}=\bar{\psi}_{0,1}+
\bar{\psi}_{1,0},~~~
\tilde{\bar{\psi}}_{0,-1}+\tilde{\bar{\psi}}_{-1,0}=\bar{\psi}_{0,-1}+
\bar{\psi}_{-1,0},
\end{align}
due to the fact: $W=r_2+r_6=r_4+r_5$. Summing up 
(\ref{psum2}) and (\ref{psum3}) gives us the desired symmetry of $Z_L$ in the interchange of
$z_i$ and $z_{i+1}$.

Similarly we prove that $Z_L(1/z_1,..)=Z_L(z_1,..)$ using the left boundary
$q$KZ equation (\ref{B$q$KZl}). Summing the three equations for $n_1=1,0,-1$ for
$\bar{\psi}_{n_1}=\sum_{\alpha}\psi_{n_1,\alpha}$ gives:
\begin{align}\label{bpsum}
U_l(\tilde{\bar{\psi_{1}}}+\tilde{\bar{\psi_{0}}}+\tilde{\bar{\psi_{-1}}})=
(k_{2,l}+k_{4,l}+k_{5,l})(\bar{\psi}_{1}+
\bar{\psi}_{-1})+(k_{1,l}+k_{3,l})\bar{\psi}_{0},
\end{align}
where now $\tilde{\bar{\psi_{1}}}=\bar{\psi}_{1}(1/z_1,..)$. Noticing again that
$U_l=k_{1,l}+k_{3,l}=k_{2,l}+k_{4,l}+k_{5,l}$
finishes the argument.

In the course of the computation of the components of $\Psi_L$ we observed that the boundary
spectral parameters appear in the components $\psi_{\pi}$ in a similar way as the
bulk spectral parameters. In particular, the $\psi_{e}$ element is symmetric in the full
set of parameters $\{x_l^{\pm 1},z_1^{\pm 1},..,z_{L}^{\pm 1},x_r^{\pm 1}\}$ and the
recurrence (\ref{rec1}) can be also applied to $x_l$ and $x_r$. The proof of this 
statement can be found in \cite{FN}. 
 
Finally, the recurrence relation (\ref{rec1}) has the initial condition $Z_0=1$. The function $Z_L$ is a polynomial in $z_i$ up to a trivial denominator which has the partial degree equal to $2L$. This follows from the formula for the fully nested element (the component $\psi_{n_1,..,n_L}$ with all $n_i=-1$) found in \cite{GN}. 
Without the loss of generality we set $z_1=w$ and consider $Z_L$ as a polynomial $Z_L(w)$ of degree $2L$. 
We find that the recurrence (\ref{rec1}) fixes the values of the polynomial $Z_L(w)$ at $2L+2$
points, i.e. when $w= z_i/\omega^2$ and $w= \omega^2/z_i$ for $i\neq 1$. This fixes $Z_L$ uniquely by the polynomial interpolation formula.

\begin{figure}[htb]
\centering
\includegraphics[width=0.3\textwidth]{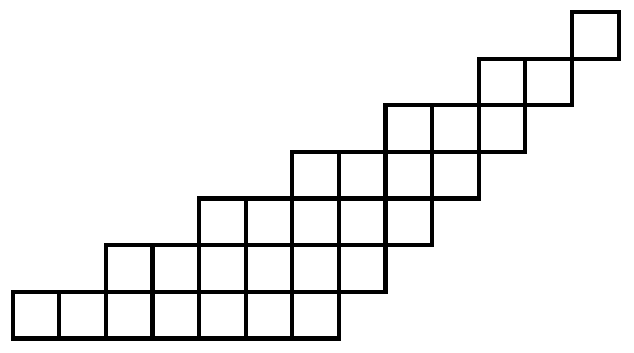}
\caption{The skew diagram for the partition $\lambda/\mu$, where $\lambda=\{13,12,11,10,9,8,7\}$ 
and $\mu=\{12,10,8,6,4,2,0\}$.}
\label{skewp}
\end{figure}

Now let us briefly mention the recurrence relation and its solution for the periodic model.
It was found in \cite{PDF} that the sum of the ground state components of the dTL model
on a cylinder of circumference $L$ satisfies:
\begin{align}\label{rec1p}
Z^p_L(z_1,..,z_{L-1}=z_{L-1}\omega,z_L=z_{L-1}/\omega)=
F_{L-1}^p(z_1,..,z_{L-1})Z^p_{L-1}(z_1,..,z_{L-1}),
\end{align}
with $F^p$:
\begin{align}\label{Ffactorp}
F^p_i(z_1,..,z_L)=z_i\prod_{1\leq j\neq i\leq L}(z_i+z_j).
\end{align}
The polynomial $Z_L^p$ is symmetric under the interchange of the rapidities but not under
their inversion. It has the initial condition $Z^p_1=1$, and is the unique solution of
(\ref{rec1p}). The form (\ref{Ffactorp}) of the recurrence factor $F^p$ suggests that
a good basis to express the solution of $Z_L^p$ is the set of elementary symmetric polynomials
for which $F^p$ is the generating function. These polynomials are defined as follows
\begin{align}\label{esp}
&E_{m}(z_{1},..,z_{L})=\sum_{1\leq i_{1}<\dots<i_{m}\leq L}z_{i_{1}}..z_{i_{m}}, \\
&E_{m}(z_{1},..,z_{L})=0~~~ \text{for}~~~ m<0,~~~ \text{and}~~~ m>L, \nonumber \\
&\prod _{i=1}^L \left(x+z_i\right)=\sum_{j=0}^L x^{L-j} E_j(z_{1},..,z_{L}). \nonumber
\end{align}
The determinant
\begin{align}\label{detesp}
Z_L^p=\det_{1\leq i,j\leq L-1} E_{3j-2i}(z_1,..,z_L)
\end{align}
solves the recurrence (\ref{rec1p}) \cite{PDF}. 
Recall the Jacobi--Trudi identity for the skew Schur polynomial $S_{\lambda/\mu}$ \cite{Mc}, 
where $\lambda$ and $\mu$ are two partitions of length $|\lambda|$ and $|\mu|$ such that each part $\mu_i$ of $\mu$ is not 
greater than the part $\lambda_i$ of $\lambda$
\begin{align}\label{skew}
S_{\lambda/\mu}=\det_{1\leq i,j \leq |\lambda|} E_{\lambda^{\prime}_i-\mu^{\prime}_j-i+j}.
\end{align}
The primed partitions are the transposed partitions of the unprimed. If we choose
$\lambda^{\prime}_i=2L-i$ and $\mu^{\prime}_j=2L-2j$ then we get (\ref{detesp}). Such a skew 
partition corresponds to the Young diagram that looks like a staircase, see Fig. \ref{skewp}.
One can as well write the formula (\ref{detesp}) for $Z_L^p$ in terms of the homogeneous 
symmetric functions.

Let us get back to (\ref{recZ}). By the analogy with the dense TL model \cite{dGPS,Ca}
the solution to this recurrence relation should involve the symplectic 
version of $S_{\lambda/\mu}$. However, to the authors knowledge the discussion of the 
skew symplectic Schur functions is missing from the literature.
The form of the $F_i$ function suggests, in turn, that a good basis for $Z_L$ is the set
of the elementary symmetric polynomials with extended list of arguments, which includes 
the symmetry in the inversion of the rapidities, i.e.:
\begin{align}\label{esp2}
\varepsilon_{m}(z_{1},..,z_{L})=E_{m}(z_{1},..,z_{L},1/z_1,..,1/z_L).
\end{align}
These symmetric polynomials are related to the elementary symmetric polynomials of $z_i$'s
and their inverses separately through the formula:
\begin{align}\label{esp2esp}
\varepsilon_{m}(z_{1},..,z_{L})=\sum_{n=0}^L E_{L-n}(z_{1},..,z_{L})
E_{L+n-m}(1/z_1,..,1/z_L).
\end{align}
The polynomial $Z_L$ can be written as a determinant of a simple matrix of
$\varepsilon_{m}$'s divided by a certain symmetric polynomial. Let us derive this formula. First, consider
\begin{align}\label{Ztilde}
\tilde{Z}_L^p(z_1,..,z_L)=Z_{2L}^{p}(z_1,..,z_L,1/z_1,..,1/z_L)=
\det_{1\leq i,j\leq 2L-1} \varepsilon_{3j-2i}(z_1,..,z_L).
\end{align}
It satisfies the following recurrence relation
\begin{align}\label{rec1pp}
\tilde{Z}^p_L(z_1,..,z_{L-1}=z\omega,z_L=z/\omega)=(z+\frac{1}{z})(z^2+1+\frac{1}{z^2}) \nonumber \\
\prod_{1\leq j\leq L-2}
\frac{(z+z_j)^2(z z_j+1)^2}{z^2 z_j^2}\tilde{Z}^p_{L-1}(z_1,..,z_{L-2},z),
\end{align}
which can be derived using (\ref{rec1p}).
The $(2L-1)\times (2L-1)$ matrix $\varepsilon_{3j-2i}$ is centrosymmetric. Indeed, a
centrosymmetric matrix of size $L\times L$ by definition is a matrix with the following 
symmetry
\begin{align}\label{cs}
M_{j,i}=M_{L-j+1,L-i+1},
\end{align}
For example, when $L=4$, $M$ has the following entries:
\begin{align}
\left(
\begin{array}{cccc}
 m_{1,1} & m_{1,2} & m_{1,3} & m_{1,4} \\
 m_{2,1} & m_{2,2} & m_{2,3} & m_{2,4} \\
 m_{2,4} & m_{2,3} & m_{2,2} & m_{2,1} \\
 m_{1,4} & m_{1,3} & m_{1,2} & m_{1,1} \\
\end{array}
\right)\nonumber
\end{align}
In our case this symmetry translates into
\begin{align}\label{epscs}
\varepsilon_{3j-2i}=\varepsilon_{3(\left(2L-1\right)-j+1)-2(\left(2L-1\right)-i+1)}=\varepsilon_{2L-1-(3j-2 i)}.
\end{align}
In order to see that the above relation indeed holds one can look at the generating function of $\varepsilon_m$
\begin{align}\label{genvar}
F_{L+1}(z_1,..,z_{L},z_{L+1}=t)=\sum_{i=0}^{2L}t^{i-L}
\varepsilon_{i}(z_1,..,z_L).
\end{align}
This function is invariant under $t\rightarrow t^{-1}$. Replacing $t\rightarrow t^{-1}$  in the right hand side in (\ref{genvar}) and changing the order of the summation one finds $\varepsilon_{m}(z_1,..,z_L)=\varepsilon_{L-m}(z_1,..,z_L)$, hence (\ref{epscs}) holds. 
A centrosymmetric matrix can be block diagonalized by the following transformation:
\begin{eqnarray}\label{TJ}
 T=
 \begin{pmatrix}
  -I & J \\
   I & J \\
 \end{pmatrix}
~~~\text{and}~~~
 T^{-1} =\frac{1}{2}
 \begin{pmatrix}
  -I &  I \\
   J &  J \\
 \end{pmatrix},
\end{eqnarray}
where, for even sizes $L=2l$, $I$ is the $l\times l$ unit matrix and $J$ is the $l\times l$ matrix with 
elements equal to $1$ on the counterdiagonal and all other elements equal to zero.
For example, applying this transformation to a centrosymmetric matrix of size $4$ gives: 
\begin{eqnarray}
T M T^{-1}=\left(
\begin{array}{cccc}
 m_{1,1}-m_{1,4} & m_{1,2}-m_{1,3} & 0 & 0 \\
 m_{2,1}-m_{2,4} & m_{2,2}-m_{2,3} & 0 & 0 \\
 0 & 0 & m_{1,1}+m_{1,4} & m_{1,2}+m_{1,3} \\
 0 & 0 & m_{2,1}+m_{2,4} & m_{2,2}+m_{2,3} \\
\end{array}
\right)\nonumber
\end{eqnarray}
For odd-size matrices $2l+1\times 2l+1$, the transformation $T$ has the same form wit the exception that it has zeros in the row $l+1$ and column $l+1$ except from the entry $(l+1,l+1)$ where it is equal to $1$. It is convenient, however, to work with the even matrices. We can rewrite (\ref{Ztilde}) as 
\begin{align*}
\tilde{Z}_L^p(z_1,..,z_L)=
\varepsilon_{2L-1}(z_1,..,z_L)\det_{1\leq i,j\leq 2L-2} \varepsilon_{3j-2i}(z_1,..,z_L).
\end{align*}
Let us  apply the transformation $T$ to the matrix the determinant above
\begin{eqnarray}\label{Tdet}
\det_{1\leq k,l\leq 2L-2} T_{k,i}\varepsilon_{3j-2i}T^{-1}_{j,l}=\frac{1}{2}
\det_{1\leq k,l\leq L-1}
(\varepsilon_{3j-2i}-\varepsilon_{3j+2i-4L})\det_{L\leq i,j\leq 2L-2}
(\varepsilon_{3j-2i}+\varepsilon_{3j+2i-4L}).
\end{eqnarray}
Wich means that $\tilde{Z}^p_L$ factorizes into two determinants and a factor $\varepsilon_{2L-1}$. These determinants define two symmetric polynomials:
\begin{eqnarray}
&&V_{L}(z_{1},..,z_{L})=\det_{1\leq i,j\leq L-1}(\varepsilon_{3j-2i}-\varepsilon_{3j+2i-4L}),
\label{V}\\
&&W_{L}(z_{1},..,z_{L})=\frac{1}{2}\det_{L\leq i,j\leq 2L-2}(\varepsilon_{3j-2i}
+\varepsilon_{3j+2i-4L}). \label{W}
\end{eqnarray}
From these formulae one can compute the degrees of these polynomials at size $L$ and $L-1$. Using this and the recurrecence relation for $\tilde{Z}_L^p$ (\ref{rec1pp}) one finds
\begin{align}
&V_{L}(z\omega,z/\omega,z_{3},..,z_{L})=\prod_{3\leq i\leq L}
\frac{(z_{i}+z)(z z_{i}+1)}{z_{i}z}(z+\frac{1}{z})V_{L-1}(z,z_{3},..,z_{L}),
\label{Vrec}\\
&W_{L}(z\omega,z/\omega,z_{3},..,z_{L})=\prod_{3\leq i\leq L}
\frac{(z_{i}+z)(z z_{i}+1)}{z_{i}z}(z^{2}+1+\frac{1}{z^{2}})
W_{L-1}(z,z_{3},..,z_{L}).\label{Wrec}
\end{align}
Alternatively, one could prove that (\ref{V}) satisfies (\ref{Vrec}) and (\ref{W})
satisfies (\ref{Wrec}) using an appropriate row-column manipulation in the corresponding
matrices. We will use this method in the next chapter to prove a similar statement for
other determinants and recurrence relations.

Let us look at  (\ref{Vrec}). Once the initial conditions are set, there is a
unique polynomial that solves this recurrence relation. If the initial condition is
$V_2(x,y)=\varepsilon_1(x,y)$, then the solution is precisely the polynomial $V_L$
defined by (\ref{V}). On the other hand,
if we take $Z_L$ and multiply it by a polynomial $P^p_L$ that satisfies:
\begin{eqnarray}\label{Pmrec}
P^p_{L}(z\omega,z/\omega,z_{3},..,z_{L})=(z+\frac{1}{z})P^p_{L-1}(z,z_{3},..,z_{L}).
\end{eqnarray}
and has the appropriate initial condition, i.e. $P^p_2(x,y)=\varepsilon_1(x,y)$ which coincides
with $V_2$, then the product
$Z_L(z_1,..,z_L)P^p_L(z_1,..,z_L)$ also satisfies the recurrence relation (\ref{Vrec}).
By the uniqueness of the solution of the recurrence  relation (\ref{Vrec})
this means $P^p_L$ divides $V_L$ and we obtain the sum rule $Z_L$:
\begin{eqnarray}\label{ZS1}
Z_{L}(z_{1},..,z_{L})=\frac{\det_{1\leq i,j\leq L-1}
(\varepsilon_{3j-2i}-\varepsilon_{3j+2i-4L})}{P^p_{L}(z_{1},..,z_{L})},
\end{eqnarray}
where $P^p_L$ can be written compactly as:
\begin{align}\label{P1}
P^p_{L}(z_{1},..,z_{L})=\frac{\I}{2(\omega-\omega^{-1})}\bigg{\{}
\prod_{j=1}^L\frac{(\omega z_j+\I)(\omega^{-1} z_j-{\I})}{z_j}-\prod_{j=1}^L
\frac{(\omega^{-1} z_j+\I)(\omega z_j-\I)}{z_j}
\bigg{\}},
\end{align}
where $\I=\sqrt{-1}$. One can now easily check the equation
(\ref{Pmrec}).
We can alternatively look for a polynomial that satisfies 
\begin{eqnarray}\label{Ppmrec}
P_{L}(z\omega,z/\omega,z_{3},..,z_{L})=(z^2+1+\frac{1}{z^2})P_{L-1}(z,z_{3},..,z_{L}),
\end{eqnarray}
and has the same initial condition as $W_L$ in (\ref{W}). Such polynomial exists:
\begin{eqnarray}\label{Pow}
P_{L}(z_{1},..,z_{L})=\frac{(-1)^L \omega}{2(1-\omega^2)(\omega-\omega^{-1})}
\bigg{\{} \prod_{j=1}^L\frac{(\omega+\omega z_j)(\omega+\omega z_j)}{z_j \omega}
\frac{(\omega+\omega^2 z_j)(\omega^2+\omega z_j)}{z_j \omega} \nonumber \\
-\prod_{j=1}^L\frac{(\omega+\omega^{-1} z_j)(\omega^{-1}+\omega z_j)}{z_j \omega}
\frac{(\omega+\omega^{-2} z_j)(\omega^{-2}+\omega z_j)}{z_j \omega}
\bigg{\}}.
\end{eqnarray}
We did not simplify the above formula in order to match it with another polynomial which will appear later. 
The product $W_L P_L$ satisfies the recurrence (\ref{Wrec}), hence: 
\begin{eqnarray}\label{ZS2}
Z_{L}(z_{1},..,z_{L})=\frac{\det_{1\leq i,j\leq L-1}
(\varepsilon_{3j-2i}+\varepsilon_{3j+2i-4L})}{P_{L}(z_{1},..,z_{L})}.
\end{eqnarray}

We close this section with the following remark. The symmetric polynomial $P^p_L$ can be written as a
ratio of two determinants: $V_{L+2}(z_1,..,z_L,\omega,\omega^2)/W_L(z_1,..,z_L)$, and since 
both $V_L$ and $W_L$ are determinants then the sum rule $Z_L$ is also a determinant of a matrix 
of the symmetric polynomials $\varepsilon_m$. This determinant is hard to write in a 
closed form, however. In
the next section we will use a different recurrence relation and hence different
basis of symmetric polynomials to express both $Z_L$ and $Z_L^p$ in determinant forms. 
Surprisingly, the second type of recurrence relations for periodic and open $Z_L$ are 
defined via a slightly more general versions of the polynomials $P^p_L$ and $P_L$.
We do not give the proofs of the recurrence relations, however, 
we believe that they can be obtained by studying the IK vertex model. The recurrence relations of the next section remain conjectural. They can be easily checked for relatively large values of the system size $L\sim 10$. 

\section{The second recurrence relation}\label{sec4}
Let us start with the sum rule $Z_L^p$ of the periodic dTL $O(1)$ ground state.
Another recurrence relation for  $Z_L^p$
appears in \cite{PDF} without a proof. This recurrence relation is unrelated to the one
we discussed previously. To write it we first extend the definition of the symmetric polynomial $P^p_L$ 
in (\ref{P1}) to:
\begin{eqnarray}\label{P}
P^p_{L}(z_{1},..,z_{L}|t)=\frac{t}{2(\omega-\omega^{-1})}\bigg{\{}
\prod_{j=1}^L(\omega z_j+t)(\omega^{-1} z_j-t)-\prod_{j=1}^L(\omega^{-1} z_j+t)(\omega z_j-t)
\bigg{\}},
\end{eqnarray}
where we omitted the $E_L$ in the denominator in  (\ref{P1}) and included 
another variable $t$. The recurrence relation reads:
\begin{align}\label{rec2p}
Z^p_L(z_1,..,z_{L-1}=t,z_L=-t)=
P^p_{L-2}(z_1,..,z_{L-2}|t)Z^p_{L-2}(z_1,..,z_{L-2}).
\end{align}
Once again we expect that a good basis of symmetric polynomials to express the solution
is given by the generating function $P^p$
\begin{align}\label{genmut}
P^p_{L}(z_1,..,z_{L}|t)=\frac{1}{2(\omega-\omega^{-1})}
\sum_{n_1,n_2=0}^L(-1)^{n_1} t^{n_1+n_2+1} E_{L-n_1}E_{L-n_2}
(\omega^{n_1-n_2}-\omega^{-n_1+n_2}).
\end{align}
Looking at the coefficients of $t$ in $P^p$ we notice that only even powers of $t$ enter this expansion. This 
can be deduced from the symmetry in the interchange of $n_1$ 
and $n_2$ in the sum in (\ref{genmut}). Hence it can be written as:
\begin{align}\label{genmu}
P^p_{L}(z_1,..,z_{L}|t)=\sum_{i=1}^{L}t^{2i}\mu_{L-i+1}(z_1,..,z_L).
\end{align}
which defines the symmetric polynomials that are quadratic in $E_m$'s:
\begin{align}\label{muesp}
\mu_{i}=\frac{1}{2(\omega-\omega^{-1})}\sum_{m=0}^{L}(-1)^{L+n}
(\omega^{2(i-m)-1}-\omega^{2(m-i)+1})E_{m}E_{2i-m-1},
\end{align}
for $i=1,..,L$, and otherwise $\mu_i=0$. We found that the solution to (\ref{rec2p})
is the following determinant written in the basis of symmetric 
polynomials $\mu_i$\footnote{This is not a basis of the space of symmetric polynomials, it is 
rather a nonlinear basis in which $Z_L^p$ can be expressed.}:
\begin{align}
&Z_L^p(z_1,..,z_L)=\det_{0\leq i,j\leq L/2-1} \mu_{3i-j+1}(z_1,..,z_L),
~~~\text{for even $L$}, \label{detmuev}\\
&Z_L^p(z_1,..,z_L)=\det_{1\leq i,j\leq (L-1)/2} \mu_{3i-j}(z_1,..,z_L),
~~~\text{for odd $L$}. \label{detmuodd}
\end{align}
Before proving this we need to examine the properties of the symmetric polynomials $\mu_i$.
In particular, how do they behave under the relevant recursion. This is best seen by
looking at the behavior of $P^p$ under the substitution $z_L=z$ and $z_{L-1}=-z$:
\begin{align}\label{recp}
P^p_{L}(z_1,..,z_{L-1}=-z,z_L=z|t)=(t^4 + t^2 z^2 + z^4)
P^p_{L-2}(z_1,..,z_{L-2}|t).
\end{align}
Comparing this to (\ref{genmu}) allows us to write the recurrence relation satisfied
by $\mu_i$'s:
\begin{align}\label{recrelmu}
\mu_{i}(z_1,..,z_{L-1}=-z,z_L=z)=z^4 \mu_{i-2}(z_1,..,z_{L-2})+
z^2 \mu_{i-1}(z_1,..,z_{L-2})+\mu_{i}(z_1,..,z_{L-2}).
\end{align}
Now we can use row column manipulations to prove, for example, that (\ref{detmuev})
satisfies (\ref{rec2p}). The proof for the odd $L$ goes in a similar manner.
First, we apply the substitution (\ref{recrelmu}) in the matrix $\mu_{3i-j+1}$, which brings
it to the form
\begin{align}\label{subm}
\tilde{\mu}_{3i-j+1}=z^4 \mu_{3i-j-1}(z_1,..,z_{L-2})+
z^2 \mu_{3i-j}(z_1,..,z_{L-2})+\mu_{3i-j+1}(z_1,..,z_{L-2}).
\end{align}
Then we subtract each column $j+1$ multiplied by $z^2$ from the column $j$ starting with
$j=1$ up to $j=L/2-2$. After this manipulation the matrix elements become:
\begin{align}\label{column}
-z^6 \mu_{3i-j-2}(z_1,..,z_{L-2})+
\mu_{3i-j+1}(z_1,..,z_{L-2}).
\end{align}
Finally, we add each row $i$ multiplied by $z^6$ to the row $i+1$, starting with $i=1$
up to $i=L/2-2$. Note, in the first row there is only one nonzero element, i.e.
$\mu_{1}$, which was unaffected by the substitution (\ref{recrelmu}), neither by the
first column manipulation since the other elements in this row are equal to zero. Therefore,
after the first row manipulation the second row is left with only one term: $\mu_{4}$.
Adding this multiplied by $z^6$ to the third row leaves it with $\mu_{7}$, and so on.
Similar subtraction happens in the other columns.
We are left then with the desired matrix $\mu_{3i-j+1}$ occupying first $L/2-2$ rows and
$L/2-2$ columns. The elements of the last row are equal to zero
except from the one that is in the last column, i.e. at the position $(L/2-1,L/2-1)$.
For convenience we will use an integer $m$ instead of $L/2$.
This matrix element is equal to the polynomial $P^p$ in the form (\ref{genmu}). The row
manipulation essentially means that we multiply each element of a row $j$ by
$z^{6 (m-1-j)}$ and then add them up starting from the top. The last element in
the column is then the sum of all the rest elements in this column thus multiplied,
so we have
\begin{align}\label{lastcolumn}
&\sum_{i=i_0}^{m-1} z^{6(m-1-i)}(\mu_{3i-m+2}+z^2\mu_{3i-m+1}+z^4\mu_{3i-m})
=z^2 \mu_{2m-2}(z_1,..,z_{2m-2}) + \nonumber \\
&+z^4 \mu_{2m-3}(z_1,..,z_{2m-2}) +z^6 (\mu_{2m-4}(z_1,..,z_{2m-2})+
z^2 \mu_{2m-5}(z_1,..,z_{2m-2})+ \nonumber \\
&+z^4 \mu_{2m-6}(z_1,..,z_{2m-2}))+ ...=P_p(z_1,..,z_{2m-2}|z),
\end{align}
where $i_0$ is the position of the first non vanishing entry on the top of the last column.
This completes the proof. 

One can alternatively view this row column manipulation 
as acting on the matrix with the entries (\ref{subm}) from the left by the matrix:
\begin{align}\label{A1}
A=\left(
\begin{array}{ccccc}
 1 & 0 & 0 & \dots & 0 \\
 z^6 & 1 & 0 & \dots & 0 \\
 z^{12} & z^6 & 1 & \dots & 0 \\
 \dots & \dots & \dots & \dots & 0 \\
 z^{6 (m-1)} & z^{6 (m-2)} & z^{6 (m-3)} & \dots & 1 \\
\end{array}
\right),
\end{align}
and also acting from the right by the matrix:
\begin{align}\label{A2}
B=\left(
\begin{array}{ccccc}
 1 & 0 & 0 & \dots & 0 \\
 -z^2 & 1 & 0 & \dots & 0 \\
 0 & -z^2 & 1 & \dots & 0 \\
 \dots & \dots & \dots & \dots & 0 \\
 0 & 0 & 0 & \dots & 1 \\
\end{array}
\right),
\end{align}
we have:
\begin{align}
\det_{0\leq k,l\leq L/2-1}A_{k,i}\tilde{\mu}_{3i-j+1}B_{j,l}=
P^p_{L-2}\det_{0\leq i,j\leq L/2-2} \mu_{3i-j+1}.
\end{align}
The introduction of these matrices in the determinant (\ref{subm}) does not affect it since  $\det A=1$ and $\det B=1$.

It seems that the determinants in (\ref{detmuev}) and (\ref{detmuodd}) must be related
to (\ref{detesp}) by some transformation, as they give the same polynomial.
Unfortunately, we are not aware of this transformation at this point. 

Let us turn to the discussion of the second recurrence relation applied to the sum rule $Z_L$ 
for the open boundary dTL model. The sum rule $Z_L$ satisfies the recurrence relation
\begin{align*}
Z_L(z_1,..,z_{L-1}=t,z_L=-t)=
P_{L-2}(z_1,..,z_{L-2}|t)Z_{L-2}(z_1,..,z_{L-2}),
\end{align*}
where the 
polynomial $P_L$ is the following
\begin{eqnarray}\label{Po}
P_{L}(z_{1},..,z_{L}|t)=\frac{(-1)^L t}{2(1-t^2)(\omega-\omega^{-1})}
\bigg{\{} \prod_{j=1}^L\frac{(t+\omega z_j)(\omega+t z_j)}{z_j t}
\frac{(t+\omega^2 z_j)(\omega^2+t z_j)}{z_j t} \nonumber \\
-\prod_{j=1}^L\frac{(t+\omega^{-1} z_j)(\omega^{-1}+t z_j)}{z_j t}
\frac{(t+\omega^{-2} z_j)(\omega^{-2}+t z_j)}{z_j t}
\bigg{\}}.
\end{eqnarray}
This polynomial itself satisfies a few recurrence relations. The one that is important for us reads
\begin{eqnarray}\label{Prec1}
P_{L}(z_{1},..,z_{L-1}=z_{L-1} \omega,z_L=z_{L-1}/\omega|t)=
(z_{L-1}^2+\frac{1}{z_{L-1}^2}-t^2-\frac{1}{t^2})P_{L-1}(z_{1},..,z_{L-1}|t).
\end{eqnarray}
If we set in this equation $t=\omega$, then it reproduces  (\ref{Ppmrec}).
Once again we consider $P_L$ as the generating function of the symmetric polynomials
which form a convenient basis to solve the recurrence for $Z_L$. First
we expand it in the elementary symmetric polynomials
\begin{align}\label{Pesp}
P_{L}(z_1,..,z_{L}|t)=\frac{(-1)^L t^{-2L}}{2(\omega-\omega^{-1})(t^2-1)}
\sum_{s=0}^{2L}t^{2s}\sum_{r=0}^{2L}(-1)^{r} (\omega^{2(r-s)+1}-\omega^{-2(r-s)-1})
\times \nonumber \\
\sum_{m,l=0}^{L} E_{L-l}(1/z_1,..,1/z_L)E_{L+l-(2s-r-1)}(z_1,..,z_L)
E_{L-m}(z_1,..,z_L)E_{L+m-r}(1/z_1,..,1/z_L),
\end{align}
which can be rewritten in terms of $\varepsilon_i$ using (\ref{esp2esp}):
\begin{align}\label{Pmu}
P_{L}(z_1,..,z_{L}|t)=\frac{(-1)^L t^{-2L}}{2(\omega-\omega^{-1})(t^2-1)}
\sum_{s=0}^{2L}t^{2s}\sum_{r=0}^{2L}(-1)^{r} (\omega^{2(r-s)+1}-\omega^{-2(r-s)-1})
\varepsilon_{r}\varepsilon_{2s-1-r}.
\end{align}
We define a new set of symmetric polynomials $\nu_i$:
\begin{align}\label{nu}
\nu_i=\sum_{j=0}^{2L}(-1)^{j+i}\varepsilon_{2i-1-j}\varepsilon_j
\frac{(\omega^{2(j-i)+1}-\omega^{-2(j-i)-1})}{2(\omega-\omega^{-1})}.
\end{align}
These symmetric polynomials are defined by the same formula as $\mu_i$ (\ref{muesp}) up to an 
overall minus sign and the replacement: $E\rightarrow \varepsilon$.

One way to find a determinant expression for $Z_L$ is to use the same 
transformation as before (\ref{TJ}) applied to (\ref{detmuev}) where $E$ is replaced 
by $\varepsilon$. Although the matrix $\mu_{3i-j+1}$
is not centrosymmetric one can, however, make it almost centrosymmetric 
by interchanging some columns. After such manipulations and a little bit of algebra 
one obtains
\begin{align}
&Z_{L}(z_1,..,z_{L})=\det_{0\leq i,j \leq m-1}
(\nu_{3i-j+1}-\nu_{3i+j+3-L}),~~~\text{for}~L=2m+1,\label{Zodd}\\
&Z_{L}(z_1,..,z_{L})=\frac{1}{P_L(z_1,..,z_L|\omega)}\det_{0\leq i,j \leq m-1}
(\nu_{3i-j+2}-\nu_{3i+j+2-L}).~~~\text{for}~L=2m.\label{Zevv}
\end{align}
This can be proven by row column manipulations similarly as before. It is not evident how 
to write (\ref{Zevv}) in a pure determinant form. 
One could try do improve it, however, we can find a nicer formula if we use different symmetric functions 
instead of $\nu_i$. Indeed, we can simply follow the idea that we used for the periodic 
cases, i.e. to use symmetric functions generated by $P_L$
\begin{align}
P_L(z_1,..,z_L|t)=\sum_{i=1}^{L-1}(t^{2i}+t^{-2i})\lambda_i+\lambda_0,
\end{align}
where $\lambda_i$ can be written in terms of $\nu_i$'s as follows
\begin{align}
\lambda_i=\sum_{k=i}^{L-1}(-1)^k\nu_{L-k}.
\end{align}
The set of polynomials $\lambda$ is a more natural basis for $Z_L$ than the set of polynomials $\nu$. 
Expressing $Z_L$ in terms of $\lambda_i$ we find a much nicer and uniform expressions for 
$Z_L$
\begin{align}
&Z_{L}(z_1,..,z_{L})=\det_{1\leq i,j \leq m}
(\lambda_{3i-j}-\lambda_{3i+j}),~~~\text{for}~L=2m+1,\label{Zodd2}\\
&Z_{L}(z_1,..,z_{L})=\det_{1\leq i,j \leq m-1}
(\lambda_{3i-j}-\lambda_{3i+j}),~~~\text{for}~L=2m.\label{Zevv2}
\end{align}
This can be proven using an appropriate row-column manipulation similarly as shown in Section \ref{sec3}.

\section{Discussions}\label{sec5}
In the dense loop model at $n=1$ \cite{PZJ,Ca,dGPS} the sum rules of 
the ground state are expressed in terms of Schur functions and symplectic Schur functions 
for periodic and open boundary conditions, respectively. 
In the case of dTL model Di Francesco found an expression for the periodic sum rule which 
is a skew Schur function (\ref{detesp}). The expression for the open boundary 
sum rule (\ref{ZS1}) or (\ref{ZS2}) reminds us of the symplectic version of the skew Schur function (\ref{detesp}) written in the form 
similar to the dual Jacobi--Trudi identity (JT). One can consult, for instance, the 
paper \cite{Fulmek_latticepath}, where many JT and dual JT identities where 
derived for the classical symmetric functions using the Gessel--Viennot algorithm \cite{GV}.
   
What is interesting about the formulae (\ref{detmuev}), (\ref{detmuodd}) is that they 
remind us the JT identities but with different symmetric polynomials. Presumably, 
there exists a version of Gessel--Viennot method that defines some symmetric functions 
by a JT-like identity, one of this symmetric functions will be equal to $Z_L^p$. 
Similarly, the skew versions of (\ref{detmuev}), (\ref{detmuodd}) appear for the open 
boundary sum rule (\ref{Zodd}), (\ref{Zevv}), and as well (\ref{Zodd2}), (\ref{Zevv2}).

We would like to emphasize two points. First, we used the prefactors in the recurrence 
relations: (\ref{Ffactorp}), (\ref{P}) and (\ref{Po}), to generate symmetric polynomials. 
These polynomials allow us to express the solutions of the corresponding recurrence 
relations in determinant forms. Instead of looking for the solutions in terms 
of Schur functions, or other symmetric functions forming a basis in the space of symmetric 
functions, one must look for the ``basis'' which is suggested by the defining equation, in our case the recurrence relations. 

The second point addresses the solutions of the recurrence relations for the open boundary 
conditions. Given a determinant expression for the periodic boundary sum rule
we assumed the second half of the variables $z_i$ ($i>m$) in the list $z_1,..,z_{2m}$ to be 
the inverses of those in the first half ($z_i,~i\leq m$) and then observed that the 
resulting matrix possesses a certain symmetry. Using an appropriate transformation (\ref{TJ})  
for this matrix allowed us to rewrite it in a block diagonal form, which means that its determinant is a 
product of the determinants of the blocks. The latter determinants turned out to be 
proportional to the open boundary sum rule (\ref{V}), (\ref{W}). 
This approach works well for both recurrence relations considered here.

Finally, we would like to stress that our results are important in the study of the correlation functions of the open boundary dTL model. 
The first progress in this direction has already been made \cite{FN}.

\section*{Acknowledgements}
A. G. was supported by the ERC grant 278124 ``Loop models, Integrability 
and Combinatorics'', AMS Amsterdam Scholarship and the Australian Research Council Centre of Excellence for Mathematical and Statistical Frontiers. A. G. thanks the University 
of Amsterdam for hospitality and support and P. Zinn-Justin and G. Feh\'{e}r for useful discussions.

\small{\bibliographystyle{plain}}
\bibliography{bibsumr}
\end{document}